\documentclass[
superscriptaddress,
nofootinbib,
twocolumn,
 amsmath,amssymb,
 aps,
pra,
notitlepage,
longbibliography
]{revtex4-1}

\usepackage{dcolumn}
\usepackage{bm}
\usepackage{epsfig}
\usepackage{graphicx}
\usepackage{latexsym}
\usepackage{amsfonts}
\usepackage{float} 
\usepackage{setspace}
\usepackage{graphicx}
\usepackage{amsmath}
\usepackage{mathrsfs}
\usepackage{verbatim}
\usepackage{color}
\usepackage{SIunits}
\usepackage{hyperref}
\usepackage{physics}

\newcommand{\be}{\begin{equation}}
\newcommand{\ee}{\end{equation}}
\newcommand{\ba}{\begin{array}}
\newcommand{\ea}{\end{array}}
\newcommand{\bqa}{\begin{eqnarray}}
\newcommand{\eqa}{\end{eqnarray}}

\begin{document}

\title{Merging mechanical bound states in the continuum in high-aspect-ratio phononic crystal gratings}

\author{Hao Tong} 
\thanks{These authors contributed equally to this work.}
\affiliation{Holonyak Micro and Nanotechnology Laboratory and Department of Electrical and Computer Engineering, University of Illinois at Urbana-Champaign, Urbana, IL 61801 USA}
\affiliation{Illinois Quantum Information Science and Technology Center, University of Illinois at Urbana-Champaign, Urbana, IL 61801 USA}
\author{Shengyan Liu} 
\thanks{These authors contributed equally to this work.}
\affiliation{Holonyak Micro and Nanotechnology Laboratory and Department of Electrical and Computer Engineering, University of Illinois at Urbana-Champaign, Urbana, IL 61801 USA}
\affiliation{Illinois Quantum Information Science and Technology Center, University of Illinois at Urbana-Champaign, Urbana, IL 61801 USA}
\author{Kejie Fang} 
\email{kfang3@illinois.edu}
\affiliation{Holonyak Micro and Nanotechnology Laboratory and Department of Electrical and Computer Engineering, University of Illinois at Urbana-Champaign, Urbana, IL 61801 USA}
\affiliation{Illinois Quantum Information Science and Technology Center, University of Illinois at Urbana-Champaign, Urbana, IL 61801 USA}

\begin{abstract} 

Mechanical bound states in the continuum (BICs) present an alternative avenue for developing high-frequency, high-$Q$ mechanical resonators, distinct from the conventional band structure engineering method. While symmetry-protected mechanical BICs have been realized in phononic crystals, the observation of accidental mechanical BICs---whose existence is independent of mode symmetry and tunable by structural parameters---has remained elusive. This challenge is primarily attributed to the additional radiation channel introduced by the longitudinal component of elastic waves. Here, we employ a coupled wave theory to predict and experimentally demonstrate mechanical accidental BICs within a high-aspect-ratio gallium arsenide phononic crystal grating. We observe the merging process of accidental BICs with symmetry-protected BICs, resulting in reduced acoustic radiation losses compared to isolated BICs. This finding opens up new possibilities for phonon trapping using BIC-based systems, with potential applications in sensing, transduction, and quantum measurements.

\end{abstract}

\maketitle

Bound states in the continuum (BICs) constitute a unique category of states existing within the continuum spectrum of open systems, yet effectively decoupled from radiating waves that could dissipate energy. BICs have been extensively  studied across various physical domains, including optics \cite{hsu2013observation,plotnik2011experimental,jin2019topologically}, acoustics \cite{cumpsty1971excitation,hein2012trapped,lyapina2015bound}, and mechanics \cite{tong2020observation,yu2022observation}. 
The distinctive features of BIC modes, including high quality factors and substantial mode sizes, have found applications in low-threshold lasing \cite{kodigala2017lasing,ren2022low,hwang2021ultralow}, ultrasensitive sensing \cite{yesilkoy2019ultrasensitive,romano2018label,wang2021ultrasensitive}, and efficient harmonic generation \cite{koshelev2019nonlinear,carletti2018giant,liu2019high}.
The introduction of the BIC concept to mechanical systems has opened a new avenue for the creation of high-quality factor, high-frequency mechanical resonators, departing from the conventional approach of band engineering with suspended structures. Mechanical BICs can be realized due to symmetry mismatch between the mechanical mode and the outgoing radiation field \cite{tong2020observation, liu2022optomechanical}---termed symmetry-protected BICs---and destructive interference between two coexisting modes \cite{yu2022observation}, known as Friedrich-Wintgen BICs. Notably, recent developments have enabled coupled mechanical BICs with co-localized optical modes in optomechanical systems \cite{liu2022optomechanical,yu2022observation}.
Compared to released mechanical micro-resonators, BIC-based mechanical systems offer advantages such as enhanced thermal capacity and macroscopic mode sizes. These characteristics hold potential for mitigating thermal noises in quantum measurements \cite{o2010quantum,palomaki2013entangling,riedinger2016non,arrangoiz2019resolving} and enabling high-throughput sensing \cite{teufel2009nanomechanical,kolkowitz2012coherent,yue2008label}.

However, a crucial breed of BICs, referred to as accidental BICs, has yet to be realized in mechanical systems. Unlike their counterparts, accidental BICs do not necessarily require radiation-forbidden symmetry or depend on the interference of two coupled modes. Instead, they emerge ``accidentally" under specific system parameters that permit zero-radiation solutions \cite{hsu2013observation}. Achieving accidental BICs in mechanical systems poses a substantial challenge, primarily due to an additional loss channel introduced by the longitudinal component of elastic waves \cite{zhao2019mechanical, tong2020observation}, in contrast to electromagnetic waves. For a mechanical resonance to manifest as an accidental BIC, it must simultaneously decouple from both transverse and longitudinal radiative waves.
Because of the distinct mechanism, accidental BICs can exist in proximity to symmetry-protected BICs by tuning system parameters, resulting in suppressed radiation loss for all surrounding modes. This merging-BIC mechanism has been shown to effectively mitigate radiation losses in structures affected by disorder-induced intermodal scattering \cite{jin2019topologically,chen2022observation,kang2022merging,hwang2021ultralow}.

Here we employ a coupled wave theory to predict and experimentally realize mechanical accidental BICs within a high-aspect-ratio gallium arsenide (GaAs) phononic crystal gratings. Additionally, the Love-wave accidental BIC can merge with the symmetry-protected BIC in the same acoustic band through single-parameter tuning. We observe the BIC merging process, resulting in enhanced mechanical quality factor of the merged BIC compared with isolated BICs. 
The demonstrated Love-wave BICs hold great promise for various applications. Particularly, Love waves offer advantages over Rayleigh waves in sensing as they effectively decouple from compressional waves in liquids \cite{jakoby1997properties,du1996study,schlensog2004love}. The BIC structure eliminates the need for multi-layered materials and complicated fabrication processes, potentially offering an improved method for utilizing Love waves in sensing applications. Moreover, the BIC phononic crystal gratings, featuring a large mode volume and a high $fQ-$product, provide a promising platform for exploring macroscopic quantum mechanical oscillators \cite{aspelmeyer2014cavity,gut2020stationary,wilson2015measurement}.

\begin{figure*}
\centering
\includegraphics[width =\linewidth]{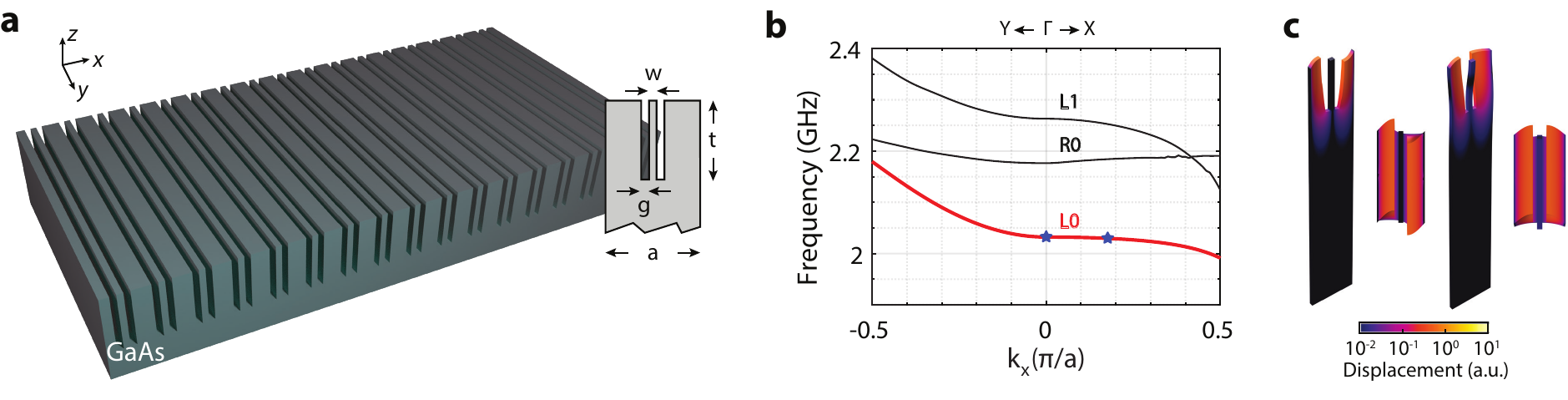}
\caption{\textbf{Phononic crystal grating with accidental mechanical BICs.} \textbf{a}. An illustration of the GaAs phononic crystal grating and the unit cell.  \textbf{b}. Mechanical band structure near the $\Gamma$ point for the unit cell dimensions indicated in the text. $L$ and $R$ indicate Love wave and Rayleigh wave, respectively. Stars indicate the symmetry-protected BIC at the $\Gamma$ point and the accidental BIC along $\Gamma-X$.  \textbf{c}. Side and top view of the mode profile of the symmetry-protected BIC (left) and the accidental BIC (right).  }
\label{fig:1}
\end{figure*}

\noindent\textbf{Results} \\
\noindent\textbf{Phononic crystal gratings with accidental mechanical BICs} The GaAs phononic crystal grating with a double-slot unit cell is illustrated in Fig. \ref{fig:1}a. For a one-dimensional phononic crystal grating aligned with the crystal axis of GaAs, the mechanical modes propagating along the $x$ direction can be divided into two categories: Love-wave modes and Rayleigh-wave modes.  The Love-wave mode is $xz-$plane-odd and thus only has $y-$displacement, while the Rayleigh-wave mode is $xz-$plane-even and thus only has coupled $x-$ and $z-$displacements. For the Rayleigh-wave mode, regardless of its symmetry with respect to the $yz-$plane, it couples with radiation fields, because the $x-$polarized plane wave is odd and the $z-$polarized plane wave is even with respect to the $yz-$plane. For the Love-wave mode, which only has the $y-$displacement, it decouples from the radiation field if it is odd with respect to the $yz-$plane, because the $y-$polarized plane wave is even. Based on the symmetry analysis, the 1D mechanical grating only supports symmetry-protected BICs that are odd Love-wave modes.  Table \ref{tab:1} lists common piezoelectric materials and their compatibility with symmetry-protected BICs in 1D phononic crystal gratings. We choose GaAs because it also has a shear piezoelectric component for excitation of the Love-wave modes. 

\begin{widetext}
\renewcommand{\arraystretch}{1.5}
\begin{table*}[!htbp]
	\centering
	\caption{\textbf{Elastic and piezoelectric properties of common piezoelectric materials. }}
	\small
	\begin{tabular}{|c|c|c|c|}
		\hline
		 Material    & GaAs   & AlN/GaN   & LiNbO$_3$  \\
		\hline
		Crystal symmetry & Cubic  & Hexagonal  & Trigonal \\
		\hline
		Mirror symmetry with respect to $xz-$ and $yz-$planes    & Yes & Yes      & No\\
		\hline
		Shear piezoelectric component  & $e_{14}=-0.16\,\mathrm{C/m^2}$ \cite{bright1989bleustein}    & None    & $e_{16}=-2.53\,\mathrm{C/m^2}$ \cite{cheng2012nanoindentation}\\
		\hline
	\end{tabular}%
	\label{tab:1}
\end{table*}
\end{widetext}

\begin{figure*}
\centering
\includegraphics[width =\linewidth]{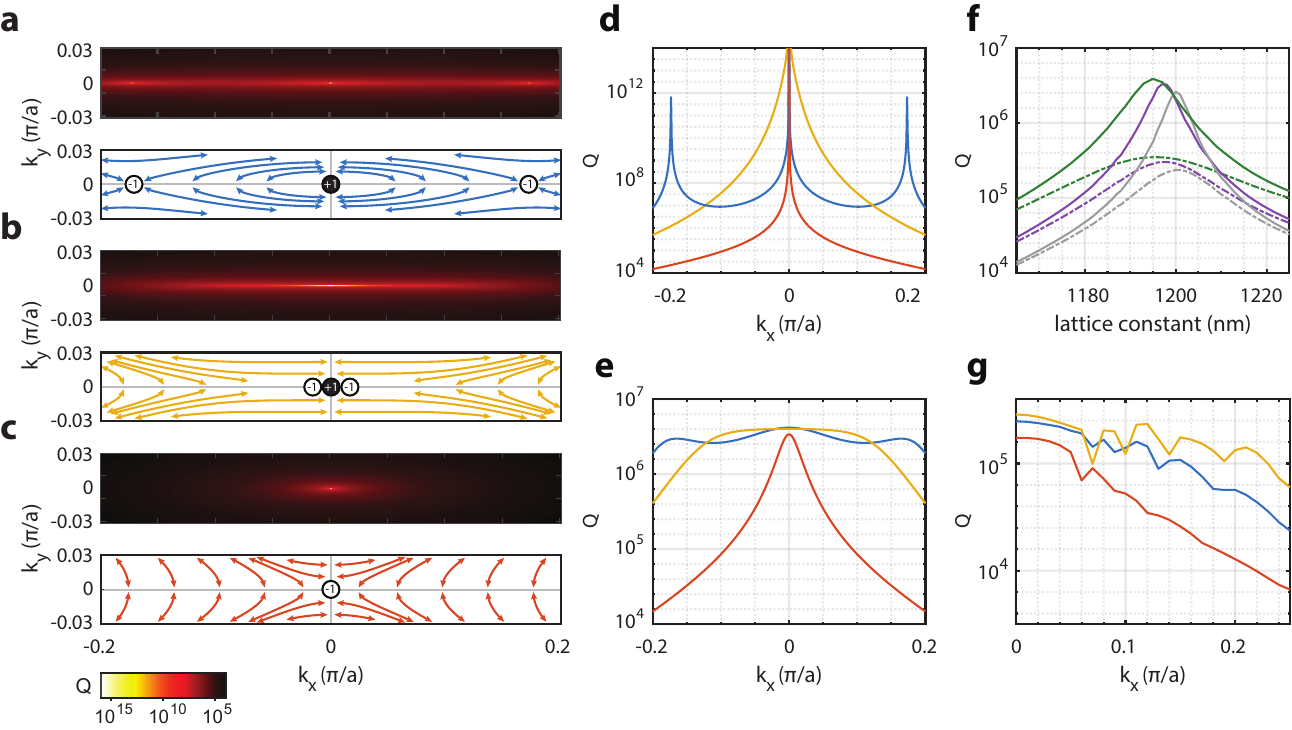}
\caption{\textbf{Merging BIC and $Q$ enhancement.} \textbf{a-c}. Numerically simulated $Q$ of the mechanical Bloch mode (top panel) and far-field polarization map (bottom panel) corresponding to before merging (\textbf{a}, $a =1198\,\mathrm{nm}$), at merging (\textbf{b}, $a =1193\,\mathrm{nm}$), and after merging (\textbf{c}, $a =1165\,\mathrm{nm}$). The topological charge of the mechanical BICs is indicated. \textbf{d} and \textbf{e}. Simulated $Q$ corresponding to $k_y=0$ (\textbf{d}) and $k_y=\pi/400$ $\upmu$m$^{-1}$ (\textbf{e}) before (blue), at (yellow), and after (red) BIC merging. \textbf{f}.  Simulated $Q$ versus lattice constant for the grating length $L=400\,\upmu$m (solid lines) and $L=120\,\upmu$m (dashed lines) and different momentum $k_x=0.09\pi/a$ (green), $k_x=0.15\pi/a$ (purple), and $k_x=0.2\pi/a$ (gray). \textbf{g}. Simulated $Q$ of $1\times 4$ supercells with $\sigma=1\%$ random variations of the slot width for lattice constant $a = 1193 \,\mathrm{nm}$ (yellow), $1185\,\mathrm{nm}$ (blue), and $1165\,\mathrm{nm}$ (red). $k_y=\pi/400$ $\upmu$m$^{-1}$. }
\label{fig:2}
\end{figure*}

We develop a coupled wave theory \cite{liang2011three,yang2014analytical,jin2019topologically} for elastic waves to identify mechanical grating structures with accidental BICs (SI). The displacement field of the Bloch mode of the phononic crystal grating can be written as ${\bm{Q}_{\bm{k}}}(\bm{r}_\parallel,z) = \sum\limits_{\bm{G}} {\bm{Q}^{\bm{G}}(z)\exp \left[ { - i(\bm{k} + \bm{G}) \cdot {\bm{r}_\parallel }} \right]}$, where ${{\bm{r}_\parallel }}$ is the in-plane spatial vector, $\bm{G}=2\pi m\bm{e_x}/a$ ($m\in \mathbf{Z}$) is the reciprocal vector and $\bm{Q}^{\bm{G}}(z)$ is the corresponding Fourier component. For Bloch modes in the vicinity of the 2nd $\Gamma$ point, i.e., $(\pm\frac{2\pi}{a},0)$, or equivalently on the 1st folded band near the $\Gamma$ point $(0, 0)$, $\bm{Q}^{\pm2\pi/a}(z)$ are the dominant Fourier components and $\bm{Q}^0(z)$ is the only radiating component \cite{zhen2014topological,jin2019topologically,yang2014analytical,liang2011three}.  According to the acoustic coupled wave theory, the radiation amplitude $\bm{Q}^0(z)$ of the Love-wave mode at the 2nd $\Gamma$ point can be calculated by, to the leading order, 
\begin{widetext} 
\be\label{cwt}
\bm{Q}^0(z) = \sum\limits_{\bm{G} =(\pm\frac{2\pi}{a},0)}\int {\mathcal{G}({z},z') {\left[ {{\omega ^2}\rho ^{ - \bm{G}} + {\partial _z}(C_{44}^{ - \bm{G}}{\partial _z})} \right]\bm{Q}^{\bm{G}}(z')} dz'},
\ee
\end{widetext} 
where $\rho ^{\bm{G}}$ and $C_{44}^{\bm{G}}$ are the Fourier components of the density and elastic tensor component $C_{44}$ of the GaAs grating layer, $\omega$ is the frequency of the Bloch mode, $\mathcal{G}= {[ - {\omega ^2}{\rho ^0} - {\partial _z}(C_{44}^0{\partial _z})]^{ - 1}}$ is the Green's function, and the integral is performed in the grating layer (SI). Symmetry-protected BIC is realized for $\bm{Q}^{\bm{G}}(z)=-\bm{Q}^{-\bm{G}}(z)$, which leads to cancellation of the two terms corresponding to $\bm{G} =(\pm2\pi/a,0)$ in Eq. \ref{cwt} and thus a vanishing radiation amplitude. On the other hand, when each individual term of Eq. \ref{cwt} is zero, an accidental BIC is realized. Using Eq. \ref{cwt}, we can predict grating structures with accidental BICs, which are close to the \emph{ab initio} simulation (SI).

Guided by the coupled wave theory and using finite-element simulations (COMSOL), we designed GaAs phononic crystal gratings with both symmetry-protected BICs and accidental BICs on the same Love-wave band. Fig. \ref{fig:1}b shows the mechanical band structure of the grating with a lattice constant $a=1198\,\mathrm{nm}$, slot width $g=100\,\mathrm{nm}$, slot depth $t=1000\,\mathrm{nm}$ and center pillar width $w=120\,\mathrm{nm}$. The lowest Love-wave band ($L_0$) supports a symmetry-protected BIC at the $\Gamma$ point and accidental BICs along the $\Gamma-X$ line at $k_x=\pm 0.18\pi/a$.  The mode profiles of the BICs are shown in Fig. \ref{fig:1}c. 

 \begin{figure*}
	\centering
	\includegraphics[width =\linewidth]{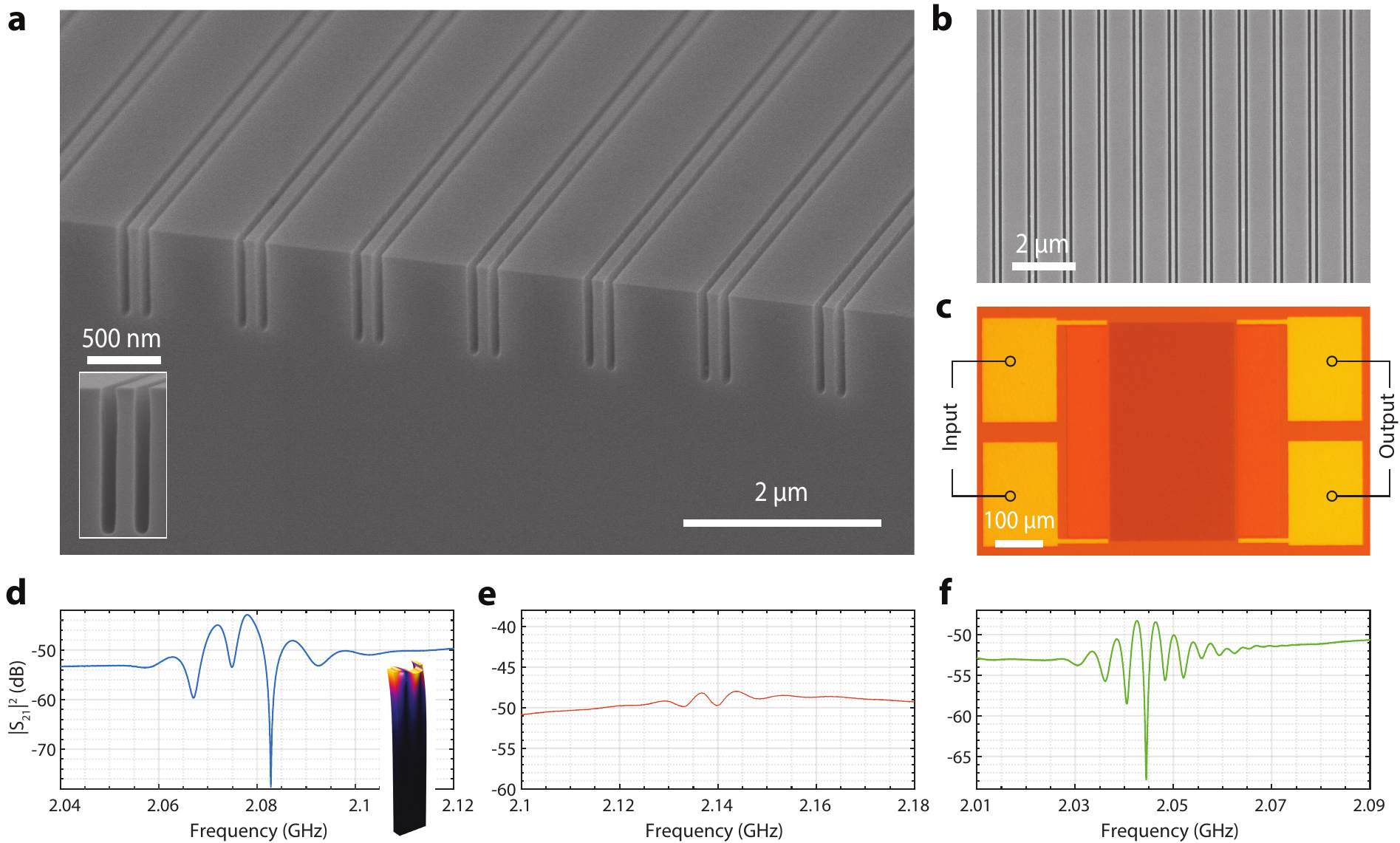}
	\caption{\textbf{Microwave transmission spectrum of the phononic crystal grating.} \textbf{a} and \textbf{b}. Scanning electron microscopy images of the GaAs phononic crystal grating. \textbf{c}. Optical microscopy image of a grating and IDTs. \textbf{d-f}. The microwave transmission spectrum of bulk GaAs (\textbf{d}), a grating with the $L_0$ band mismatched with the IDT frequency (\textbf{e}), and a grating with the $L_0$ band matched with the IDT frequency (\textbf{f}). The IDT Love-wave mode profile is shown in the inset of \textbf{d}. }
	\label{fig:3}
\end{figure*}

\noindent\textbf{Merging mechanical BICs} 
The $k-$position of the accidental BICs can be tuned by varying the grating parameters. For example, by reducing the lattice constant from 1198 nm to 1193 nm, the two accidental BICs shift towards the $\Gamma$ point and eventually merge with the symmetry-protected BIC, forming a merged BIC (Figs. \ref{fig:2}a and b). When the lattice constant is below 1193 nm, the grating structure only has one symmetry-protected BIC at the $\Gamma$ point (Fig. \ref{fig:2}c, $a = 1165$ nm).  
Such a merging behavior of the mechanical BICs is also illustrated by their topological charges. Because BICs have vanishing radiation amplitudes, the far-field polarization of BICs cannot be defined, which leads to singularity points in the far-field polarization map of Bloch modes (see Figs. \ref{fig:2}a-c bottom panel). These singularity points are characterized by a topological charge, i.e., the winding number of the polarization surrounding them \cite{zhen2014topological, tong2020observation}. During the merging process, the total topological charge is conserved \cite{zhen2014topological}. Figs. \ref{fig:2}a-c show the $x-$ and $y-$components of the far-field polarization of Bloch modes of the $L_0$ band and the topological charges of mechanical BICs before, at, and after BIC merging. On the other hand, due to the odd symmetry of the Love-wave mode, the $z-$component of the far-field polarization vanishes for Love-wave Bloch modes on the high-symmetric $k_x$ and $k_y$ axes, which are represented by nodal lines in the polarization map. 

The merging of BICs enhances the quality factor ($Q$) of the Bloch modes in the vicinity of the BICs \cite{zhen2014topological,jin2019topologically,kang2022merging}. For an isolated BIC, the scaling of $Q$ versus $k_x$ along the $\Gamma-X$ line is given by $Q\propto 1/k_x^2$ for $k_x\ll \pi/a$ (Fig. \ref{fig:2}d red line) \cite{jin2019topologically} (SI).  In the presence of two accidental BICs adjacent to the symmetry-protected BIC, the $Q$ scaling becomes $Q\propto 1/[k_x^2(k_x-k_{\mathrm{BIC}})^2(k_x+k_{\mathrm{BIC}})^2]$ (Fig. \ref{fig:2}d blue line), where $\pm k_{\mathrm{BIC}}$ are the momenta of the accidental BICs (SI). At the merging point, i.e., $k_{\mathrm{BIC}}=0$, the merged BIC possesses $Q\propto 1/k_x^6$, leading to enhanced quality factor for the Bloch modes near the merged BIC (Fig. \ref{fig:2}d yellow line).  The $Q$-enhancement effect due to the merging BIC is also manifested at finite $k_y\ll \pi/a$, despite the disappearance of rigorous BICs for finite $k_y$. Fig. \ref{fig:2}e shows the simulated $Q(k_x)$ for $k_y=\pi/400$ $\upmu$m$^{-1}$ and $a=1198, 1193, 1165$ nm. 

This result indicates the effect of merging BIC can persist in finite grating structures. The standing-wave resonance of order $\{m, n\}$ in finite grating structures is characterized by quantized momentum $(k_x, k_y)=(m\pi/Na, n\pi/L)$, where $N$ is the number of grating periods and $L$ is the length of the grating along the $y$ direction. To illustrate the effect of merging BIC on the $Q$ of standing-wave resonances, we simulate the $Q$ of Bloch modes of finite $(k_x, k_y)$ for varying lattice constant across the merging point.  In Fig. \ref{fig:2}f, solid (dashed) lines represent the $Q$ of Bloch modes with $k_y$ corresponding to $n=1$ and $L=400\,\upmu$m ($L=120\,\upmu$m). For each $k_y$, three different $k_x$ are simulated, $k_x=0.09~\pi/a$, $0.15~\pi/a$, and $0.2~\pi/a$, corresponding to green, purple, and gray lines. We find a peaked $Q$ exists for all these cases as a result of merging BICs, and the value of the peaked $Q$ becomes larger for smaller $k_x$ and $k_y$ as the Bloch mode approaches the merged BIC.

\begin{figure*}
	\centering
	\includegraphics[width =\linewidth]{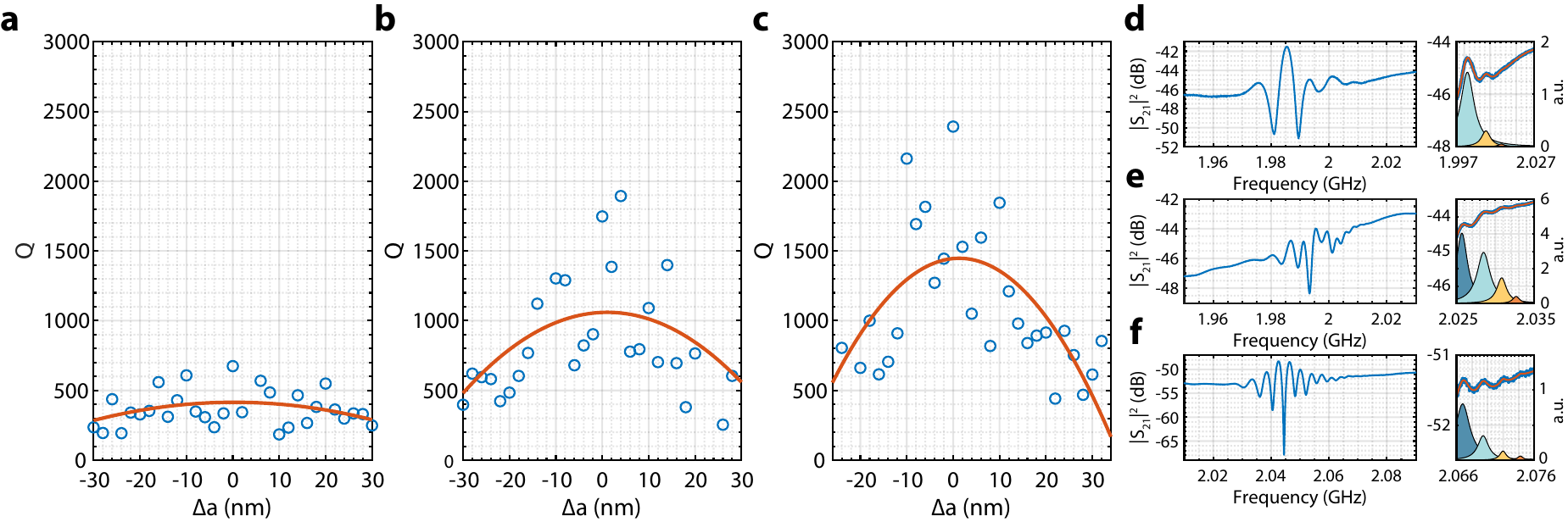}
	\caption{\textbf{Observation of merging mechanical BIC.} \textbf{a-c}. Measured $Q$ of the observed lowest-order standing-wave resonance in the three sets of gratings with varying lattice constants. \textbf{a}: ($N=50$, $L=120\,\upmu$m), \textbf{b}: ($N=100$, $L=120\,\upmu$m), \textbf{c}: ($N=200$, $L=400\,\upmu$m). Red lines are polynomial fitting of the $Q$ data. \textbf{d-f}. Microwave transmission spectrum of the gratings with the highest $Q$ in the three sets. Right panels are zoom-in plots in narrow frequency range. Red lines are model fitting. Background-subtracted, fitted resonances are also shown with the amplitude measured by the linear scale on the right axis.}
	\label{fig:4}
\end{figure*}

The enhanced quality factor due to the merging BIC is also manifested in grating structures with disorders \cite{regan2016direct,jin2019topologically}. This is because the merging BIC enhances the quality factor of all adjacent Bloch modes as shown in Figs. \ref{fig:2}d and e, thus suppressing the radiation loss due to the disorder-induced intermodal scattering \cite{regan2016direct,jin2019topologically}.  To show this, we simulate a super-cell structure with 4 unit cells and each unit cell has a random variation of the slot width (standard deviation $\approx$ 1\%). As shown in Fig. \ref{fig:2}g, for the same disorder level, the $Q$ of the grating with the merging BIC ($a = 1193\,\mathrm{nm}$) is higher than that of the grating with isolated BICs ($a = 1165\,\mathrm{nm}$). Such protection against disorders remains effective for a wide range of $k_x$ and lattice constants near the merging-BIC design, which reveals the robustness of the merging-BIC mechanism. Since disorders are inevitable in fabricated structures, the merging BIC mechanism is expected to mitigate scattering losses compared to structures with only isolated BICs.

\noindent\textbf{Experimental demonstrations} To experimentally demonstrate the accidental mechanical BIC and the BIC merging process, we fabricated GaAs phononic crystal gratings (Figs. \ref{fig:3}a and b; see Methods for device fabrication). We also fabricated interdigital transducers (IDTs) for piezoelectric excitation and probe of surface acoustic waves coupled with the phononic crystal grating (Fig. \ref{fig:3}c). A vector network analyzer is used to measure the microwave transmission of the device via RF probes that contact with the IDTs. The mechanical mode spectrum of the phononic crystal grating thus can be inferred from $|S_{21}|^2$. For the IDT with a pitch about 1600 nm, a Love-wave mode with a frequency around 2 GHz can be excited (see Fig. \ref{fig:3}d inset). A typical IDT transmission spectrum through the bulk GaAs without the phononic crystal grating is shown in Fig. \ref{fig:3}d, where the number of IDT finger pairs is 50, resulting in a bandwidth of about 60 MHz. We then used such IDTs to probe phononic crystal gratings. The higher-frequency Love-wave bands of the grating correspond to higher-order modes along the $z$ direction, which mode-mismatch with the fundamental Love-wave mode of the IDT. Consequently, the IDT transmission is significantly attenuated if the IDT frequency is higher than the $L_0$ band (along $k_x$) of the grating (Fig. \ref{fig:3}e). In contrast, when the IDT frequency lies within the $L_0$ band (along $k_x$) of the grating, the IDT transmission is strongly modulated by the grating and multiple standing-wave resonances of the grating can be observed in the spectrum (Fig. \ref{fig:3}f). 

We fabricated three sets of phononic crystal gratings with different number of grating periods $N$ and grating length $L$: ($N=50$, $L=120\,\upmu\mathrm{m}$), ($N=100$, $L=120\,\upmu\mathrm{m}$), and ($N = 200$, $L= 400 \,\upmu\mathrm{m}$). In each set, the lattice constant is scanned in a range of 60 nm around $a=1193$ nm, corresponding to the merging BIC design, with a step size of 2 nm. We measured the microwave transmission of each device at room temperature and inferred the $Q$ of the standing-wave resonances of the grating by fitting the transmission spectrum (SI). The $Q$ of the observed lowest-order standing-wave resonance in each device is summarized in Figs. \ref{fig:4}a-c for the three sets of gratings. Within each set of gratings, the measured $Q$ varies with the lattice constant and peaked at a lattice constant around $a=1193$ nm. This is a direct demonstration of the BIC merging process and is consistent with the simulation (Fig. \ref{fig:2}f). 
Comparing different sizes of gratings, the ($N = 200$, $L= 400 \,\upmu\mathrm{m}$) set has the highest $Q$ at the merging point, because its standing-wave resonances have the smallest $k_y$, which agrees well with the simulated result shown in Fig. \ref{fig:2}f. 

Figs. \ref{fig:4}d-f show the transmission spectrum of the gratings with the highest $Q$ in the three sets. The measured transmission spectrum can be fit using a coupled mode theory (SI). For the transmission coefficient at the resonance frequency, $|S_{12}|^2\propto (\frac{\kappa_e}{\kappa})^2$, where $\kappa\equiv \omega/Q$ is the total dissipation rate of the standing-wave resonance and $\kappa_e$ is the coupling rate with the IDT-excited surface acoustic wave. As a result, given the similar noise floor in the transmission spectrum and signal-to-noise ratio, the lowest-order resonances observed in the three sets of gratings have a comparable $\kappa_e/\kappa$ ratio. We examine three gratings, with one from each set, that are furthest away from the merging BIC point. The $Q$ of the lowest-order resonance of these three gratings have a ratio about $Q_{N=50}:Q_{N=100}:Q_{N=200} \approx 1:2:4$,
(i.e., $Q_{N=50}=235$, $Q_{N=100}=397$, $Q_{N=200}=805$), 
and thus we infer $Q_{e,N=50}:Q_{e,N=100}:Q_{e,N=200} \approx 1:2:4$ ($Q_e\equiv \omega/\kappa_e$). Our analysis shows that this can only be satisfied when the observed lowest-order resonances have a similar $k_x(=m\pi/Na)$, which is found to be in the range of $(0.04\pi/a, 0.08\pi/a)$ (SI). This is consistent with the lower $Q$ observed for the $N=50$ set compared to the $N=100$ set, because the resonance of a smaller grating has a broader momentum distribution than that of a larger grating with the same $k_x$ (SI), leading to more radiation losses.

In conclusion, we have realized mechanical accidental BICs and their merging with symmetry-protected BICs in GaAs phononic crystal gratings. The merging BIC is experimentally verified via the observation of suppressed acoustic radiation loss as compared with isolated BICs. The acoustic coupled wave theory developed here can be used to explore accidental BICs and BIC merging in other mechanical systems beyond 1D gratings. The observation of mechanical merging BIC enables an alternative approach for creating mechanical oscillators with high frequencies and macroscopic sizes, which hold great promises in various applications from signal transduction to sensing.

\noindent\textbf{Methods}\\
\textbf{Device fabrication.} The devices are fabricated on a GaAs wafer with $[100]$ orientation. A 140 nm thick SiO$_2$ is deposited as the hard mask. The phononic crystal gratings are patterned by electron beam lithography using ZEP 520A as the mask. The pattern is transferred to the hard mask by inductively coupled plasma reactive ion etch (ICP-RIE) of SiO$_2$ using SF$_6$ and CHF$_3$, and subsequently transferred to GaAs with another ICP-RIE using BCl$_3$, Ar and N$_2$. The IDTs are defined by a second electron beam lithography with PMMA as the mask, followed by electron beam evaporation of 10 nm chromium and 70 nm gold and subsequent lift-off process. \\ 


\vspace{2mm}
\noindent\textbf{Data availability}\\ 
Data supporting the findings of this study are available within the article and its Supplementary Information, or from the corresponding author upon reasonable request.

\noindent\textbf{Acknowledgements}\\ 
This work is supported by the U.S. National Science Foundation (Grant No. 1944728 and 2137642) and the Office of Naval Research (Grant No. N00014-21-1-2136).

\noindent\textbf{Author contributions}\\ 
S.L. developed the coupled wave theory. S.L. and H.T. designed the device. H.T. manufactured the device. H.T. and S.L. performed the measurement and data analysis. All authors contributed to the writing of the manuscript.

%
 
\end{document}


\title{Supplementary Information for ``Merging mechanical bound states in the continuum in high-aspect-ratio phononic crystal gratings''}
	
	\author{Hao Tong} 
	\thanks{These authors contributed equally to this work.}
	\affiliation{Holonyak Micro and Nanotechnology Laboratory and Department of Electrical and Computer Engineering, University of Illinois at Urbana-Champaign, Urbana, IL 61801 USA}
	\affiliation{Illinois Quantum Information Science and Technology Center, University of Illinois at Urbana-Champaign, Urbana, IL 61801 USA}
	\author{Shengyan Liu} 
	\thanks{These authors contributed equally to this work.}
	\affiliation{Holonyak Micro and Nanotechnology Laboratory and Department of Electrical and Computer Engineering, University of Illinois at Urbana-Champaign, Urbana, IL 61801 USA}
	\affiliation{Illinois Quantum Information Science and Technology Center, University of Illinois at Urbana-Champaign, Urbana, IL 61801 USA}
	\author{Kejie Fang} 
	\email{kfang3@illinois.edu}
	\affiliation{Holonyak Micro and Nanotechnology Laboratory and Department of Electrical and Computer Engineering, University of Illinois at Urbana-Champaign, Urbana, IL 61801 USA}
	\affiliation{Illinois Quantum Information Science and Technology Center, University of Illinois at Urbana-Champaign, Urbana, IL 61801 USA}
	\maketitle

\section{Scaling rule of the radiative $Q$ for the merging BIC in phononic crystal gratings} 
We can expand the radiation amplitudes, i.e., the far-field polarization, $c_x$, $c_y$, and $c_z$, of the Bloch mode in the vicinity of the $\Gamma$ point as
	\begin{equation}
		{c_x}(k) = \sum {{p_{mn}}k_x^mk_y^n}, \quad {c_y}(k) = \sum {{q_{mn}}k_x^mk_y^n}, \quad  {c_z}(k) = \sum {{r_{mn}}k_x^mk_y^n}.
	\end{equation}
	The Bloch mode $\bm{u}$ at wavevector $\bm{k}$ and $\mathcal{R}\bm{k}$ can be related by
	\begin{equation}
		{\bm{u}_{\mathcal{R}\bm{k}}} = {\alpha _\Gamma }{O_{\mathcal{R}}}{\bm{u}_{\bm{k}}},
	\end{equation}
	where $\alpha_\Gamma$ is the character of $\mathcal{R}$ at $\Gamma$ point. For the Love-wave mode considered here, because of the symmetry, we have
	\begin{equation}
		\begin{aligned}\label{c}
			{c_x}({k_x},{k_y}) &= {c_x}( - {k_x},{k_y}) = {c_x}( - {k_x}, - {k_y}),\\
			{c_y}({k_x},{k_y}) &=  - {c_x}( - {k_x},{k_y}) = {c_x}({k_x}, - {k_y}),\\
			{c_z}({k_x},{k_y}) &=  - {c_z}( - {k_x},{k_y}) =  - {c_z}({k_x}, - {k_y}).
		\end{aligned}
	\end{equation}
As a result, $c_x$, $c_y$, and $c_z$ have the following form,
	\begin{equation}
			\begin{aligned}
		{c_x}(\bm{k}) = {p_1}{k_y} + {p_2}k_y^3 + {p_3}k_x^2{k_y}+O(\bm{k}^5),\\
		{c_y}(\bm{k}) = {q_1}{k_x} + {q_2}k_x^3 + {q_3}{k_x}k_y^2+O(\bm{k}^5),\\
	{c_z}(\bm{k}) = {r_1}{k_x}{k_y} + {r_2}{k_x}k_y^3 + {r_3}k_x^3{k_y}+O(\bm{k}^5).
			\end{aligned}
	\end{equation}
The acoustic energy flux is defined as
\begin{equation}
	\bm{P} =  - \bm{\tau} {{\dot{\bm{Q}} }^*},
\end{equation}
where $\tau$ is the stress tensor and $\bm{Q}$ is the displacement. In our phononic crystal system, we have the $z$ direction flux
\begin{equation}
	{P_z} = {C_{44}}\frac{{{\omega ^2}}}{{{v_T}}}{\left| {{c_x}} \right|^2} + {C_{44}}\frac{{{\omega ^2}}}{{{v_T}}}{\left| {{c_y}} \right|^2} + {C_{11}}\frac{{{\omega ^2}}}{{{v_L}}}{\left| {{c_z}} \right|^2}
\end{equation}
for (100) orientated cubic crystal, where $C_{11}$ and $C_{44}$ are the diagonal components of the elastic tensor of GaAs and $v_T$ ($v_L$) is the transverse (longitudinal) speed of sound. Using Eq. \ref{c}, we have, for Bloch modes on the $k_x-$axis,	
\begin{equation}
	{P_z} = {C_{44}}\frac{{{\omega ^2}}}{{{v_T}}}q_1^2k_x^2.
\end{equation}
The radiative quality factor $Q$ is inverse proportional to the acoustic energy flux 
and thus
\begin{equation}
Q\propto1/k_x^2.
 \end{equation}
For the accidental BICs on the $k_x$ axis with wavevector $(\pm k_{\mathrm{BIC}},0)$, we have an extra constraint as
	\begin{equation}
		{c_y}({k_{\mathrm{BIC}}},0) =  - {c_y}( - {k_{\mathrm{BIC}}},0) = {k_{\mathrm{BIC}}}({q_1} + {q_2}k_{\mathrm{BIC}}^2) = 0,
	\end{equation}
	which yields
	\begin{equation}
		{q_1} + {q_2}k_{\mathrm{BIC}}^2 = 0
	\end{equation}
	and thus
	\begin{equation}
		{c_y}(k) = {q_2}{k_x}(k_x^2 - k_{\mathrm{BIC}}^2).
	\end{equation}
As a result, in the presence of accidental BICs, 
	\begin{equation}
		{P_z} = {C_{44}}\frac{{{\omega ^2}}}{{{v_T}}}q_2^2k_x^2{({k_x} + {k_{\mathrm{BIC}}})^2}{({k_x} - {k_{\mathrm{BIC}}})^2}
	\end{equation}	
and thus 
	 	\begin{equation}
	 	Q \propto 1/k_x^2{({k_x} + {k_{\mathrm{BIC}}})^2}{({k_x} - {k_{\mathrm{BIC}}})^2}.
	 	\end{equation}	
	
For Bloch modes with finite $k_y$, the leading-order terms in $c_x$ and $c_z$, i.e., $p_1 k_y$ and $r_1 k_xk_y$, need to be much smaller than that in $c_y$, i.e., ${q_2}{k_x}(k_x^2 - k_{\mathrm{BIC}}^2) + {q_3}{k_x}k_y^2$, for observation of the merging BIC effect, which requires a sufficiently long grating along the $y$-axis.
	
\section{Coupled-wave theory for phononic crystal gratings} 
In the section, we derive a coupled-wave theory (CWT) for Love-wave modes in the phononic crystal grating. We use the CWT to calculate the radiation amplitude of the Bloch mode and predict the accidental BIC in phononic crystal gratings.

\subsection{Eigenmode equation for Love-wave modes in phononic crystal gratings}
The elastic wave equation 
\begin{equation}
		\rho(\bm{r}) \ddot{Q}_i(\bm{r})=\partial_j\left(C_{i j k l}(\bm{r}) \partial_k Q_l(\bm{r})\right)
\end{equation}
governs the propagation of the waves in the phononic crystal. The density $\rho(\bm{r})$ and elasticity tensor $C_{i j k l}(\bm{r}) $ are periodic functions and can be expanded as $\rho(\bm{r})=\sum_{\bm{G}} \rho^{\bm{G}}(z) e^{-i \bm{G} \cdot \bm{r}_{\|}}$ and $C_{ijkl}(\bm{r}) = \sum_{\bm{G}} C_{i j k l}^{\bm{G}}(z) e^{-i \bm{G} \cdot \bm{r}_{\|}}$, where $\bm{G}$ is the reciprocal vector of the phononic crystal.  The displacement field $\bm{Q}(\bm{r}_{\|},z)$ of the Bloch mode can be decomposed as $Q_{i,\bm{k}_{\|}}=\sum_{\bm{G}} Q_i^{\bm{G}}(z) e^{i\left(\omega t-\left(\bm{k}_{\|}+\bm{G}\right) \cdot \bm{r}_{\|}\right)}$ according to Bloch's theorem.

Suppose we consider Love-wave modes with displacement along $y$ and propagating along $x$ in 1D phononic crystal gratings, the eigenmode equation for the Bloch mode is given by 
	\begin{equation}
		-\omega^2 \sum_{\boldsymbol{G}^{\prime}} \rho^{\boldsymbol{G}-\boldsymbol{G}^{\prime}}(z) Q_y^{\boldsymbol{G}^{\prime}}(z) e^{-i\left(\boldsymbol{k}_{\|}+\boldsymbol{G}\right) \cdot \boldsymbol{r}}=\sum_{\boldsymbol{G}^{\prime}} \partial_j\left(\left[C_{y j k y}^{\boldsymbol{G}-\boldsymbol{G}^{\prime}}(z) e^{-i\left(\boldsymbol{G}-\boldsymbol{G}^{\prime}\right) \cdot \boldsymbol{r}}\right]\left[\partial_k Q_y^{\boldsymbol{G}^{\prime}}(z) e^{-i\left(\boldsymbol{k}_{\|}+\boldsymbol{G}^{\prime}\right) \cdot \boldsymbol{r}}\right]\right).
	\end{equation}
We can equate all the terms with the same reciprocal vector $\bm{G}$ and obtain
	\begin{equation}
		\label{eqn:eigenp}
		\begin{aligned}
			- {\omega ^2}{\rho ^0}(z)Q_y^{\bm{G}}(z) - {\partial _z}(C_{44}^0(z){\partial _z}Q_y^{\bm{G}}(z)) + {\beta ^2}C_{66}^0(z)Q_y^{\bm{G}}(z) = \\
			\sum\limits_{{\bm{G}}' \neq \bm{G}} {{\omega ^2}{\rho ^{{\bm{G}} - {\bm{G}}'}}(z)Q_y^{{\bm{G}}'}(z) + {\partial _z}(C_{44}^{{\bm{G}} - {\bm{G}}'}(z){\partial _z}Q_y^{{\bm{G}}'}(z)) - \beta \beta 'C_{66}^{{\bm{G}} - {\bm{G}}'}(z)Q_y^{{\bm{G}}'}(z)},
		\end{aligned}
	\end{equation}
	where $\beta = k_\parallel+G$, $\beta'=k_\parallel+G'$. Eq. \ref{eqn:eigenp} applies to crystals with the orthotropic symmetry (including cubic symmetry, e.g., GaAs) and transversely isotropic symmetry.

\subsection{Solution of the first-order Fourier component}
We solve for the Fourier components $Q_y^{G=\pm 2\pi/a}$ of the Bloch mode at the 2nd $\Gamma$ point (i.e., $k_\|=0$) of the 1D grating using the coupled wave equation Eq. \ref{eqn:eigenp}. Because the Bloch mode is at the 2nd $\Gamma$ point, its dominant Fourier components are $Q_y^{G=\pm 2\pi/a}$. As a result, for $G=2\pi/a$, Eq. \ref{eqn:eigenp} becomes
		\begin{equation}
			\label{eqn:PWG}
			\begin{aligned}
				- {\omega_m ^2}{\rho ^0}(z)Q_y^{2\pi /a}(z) - {\partial _z}(C_{44}^0(z){\partial _z}(Q_y^{2\pi /a}(z)) + {(2\pi /a)^2}C_{66}^0(z)Q_y^{2\pi /a}(z)\\
				= {\omega_m ^2}{\rho ^{4\pi /a}}(z)Q_y^{ - 2\pi /a}(z) + {\partial _z}(C_{44}^{4\pi /a}(z){\partial _z}(Q_y^{ - 2\pi /a}(z)) + {(2\pi /a)^2}C_{66}^{4\pi /a}(z)Q_y^{ - 2\pi /a}(z).
			\end{aligned}
		\end{equation}
In contrast to the photonic CWT \cite{peng2011coupled}, where the R.H.S. of Eq. \ref{eqn:PWG} is ignored, these terms have to be contained in the acoustic CWT because of the strong perturbation introduced by the void in the grating.

To solve Eq. \ref{eqn:PWG} for the grating, we separate the grating structure into two layers: the grating layer and the uniform substrate layer, with the interface at $z=0$, and use the following ansatz 
	\begin{equation}\label{ansatz}
		Q_y^{2\pi/a} (z) \equiv \Theta_y(z) = \left\{ {\begin{array}{*{20}{c}}
				{B\sin {\kappa _0}z + C\cos {\kappa _0}z,}&{z \ge 0},\\
				{A{e^{{\kappa _{ - 1}}z}},}&{z < 0}.
		\end{array}} \right.
	\end{equation}
We also assume $Q_y^{ - 2\pi /a}(z) =  - Q_y^{2\pi /a}(z)$ since we consider the presence of a symmetry-protected BIC. The equation for $Q^{-2\pi/a}$ is the same as Eq. \ref{eqn:PWG} under this assumption. We denote the Fourier components in the grating layer and substrate layer with subscript $0$ and $-1$, respectively.   Substituting the ansatz of Eq. \ref{ansatz} into Eq. \ref{eqn:PWG} yields
	\begin{equation}
		\label{eqn:k0}
		\kappa _0^2 = \frac{{\rho _0^0 - \rho _0^{4\pi /a}}}{{C_{44,0}^0 - C_{44,0}^{4\pi /a}}}\omega _m^2 - \frac{{C_{66,0}^0 + C_{66,0}^{4\pi /a}}}{{C_{44,0}^0 - C_{44,0}^{4\pi /a}}}{\left( {\frac{{2\pi }}{a}} \right)^2}
	\end{equation}
	and
	\begin{equation}\label{eqn:k-1}
		\begin{aligned}
			\kappa _{ - 1}^2 = \frac{{C_{66, - 1}^0}}{{C_{44, - 1}^0}}{\left( {\frac{{2\pi }}{a}} \right)^2} - \frac{{\rho _{ - 1}^0}}{{C_{44, - 1}^0}}\omega _m^2.
		\end{aligned}
	\end{equation}
Since the top surface of the grating at $z=h$ is a free boundary, we have the boundary condition
	\begin{equation}\label{S20}
		\partial Q_y^{2\pi/a} (z)/\partial z|_{z=h}=B\cos {\kappa _0}h - C\sin {\kappa _0}h = 0.
	\end{equation} 
For the slab-substrate interface, the displacement $Q_y^{2\pi/a}$ is continuous, which leads to $A=C$. In addition, integrating Eq. \ref{eqn:PWG} from $z=0^-$ to $z=0^+$ results in the continuous condition of $C_{44}^0(z){\partial _z}Q_y^{2\pi /a}(z) + C_{44}^{4\pi /a}(z){\partial _z}Q_y^{ - 2\pi /a}(z)$, leading to
	\begin{equation}\label{S21}
		\left( {C_{44,0}^0 - C_{44,0}^{4\pi /a}} \right)B{\kappa _0} = C_{44, - 1}^0C{\kappa _{ - 1}}.
	\end{equation}
From Eqs. \ref{S20} and \ref{S21}, we obtain
	\begin{equation}\label{eqn:dis} 
		\tan {\kappa _0}h = \frac{C_{44, - 1}^0{\kappa _{ - 1}}}{{(C_{44,0}^0 - C_{44,0}^{4\pi /a}} ){\kappa _0} }.
	\end{equation}
Eqs. \ref{eqn:k0}, \ref{eqn:k-1} and \ref{eqn:dis} together can determine $\omega_m$, $\kappa_0$, and $\kappa_{-1}$ for a given grating structure.

\subsection{Green's function of zeroth-order Fourier component}
In the previous section, we obtained the first-order Fourier component of the Bloch mode at the 2nd $\Gamma$ point, which can be used to find the zeroth-order Fourier component ($\bm G=0$) using the coupled wave equation Eq. \ref{eqn:eigenp}. To this end, we will use the approach of the Green's function to solve Eq. \ref{eqn:eigenp} for the zeroth-order Fourier component. In this section, we first find the Green's function corresponding to Eq. \ref{eqn:eigenp}. 

The Green's function $\mathcal{G}(z,z')$ corresponding to Eq. \ref{eqn:eigenp} satisfies
		\begin{equation}
			\label{eqn:Gp}
			- {\omega ^2}{\rho ^0}(z)\mathcal{G}(z,z') - {\partial _z}(C_{44}^0(z){\partial _z}(\mathcal{G}(z,z'))) + \beta ^2C_{66}^0(z)\mathcal{G}(z,z') = \delta (z-z')
		\end{equation}
We assume $z' \in (0,h)$. For $z>z'$, the Green's function can be taken as 
	\begin{equation}
		\label{eqn:>G}
		\mathcal{G}(z,z') = \begin{array}{*{20}{c}}
			{{J_0}{e^{ - i{k_{0,z}}(z - z')}} + {K_0}{e^{i{k_{0,z}}(z - z')}}}&{z \le h}
		\end{array}
	\end{equation}
	and for $z<z'$, we take the Green's function as 
	\begin{equation}
		\label{eqn:<G}
		\mathcal{G}(z,z') = \left\{ {\begin{array}{*{20}{c}}
				{{N_{ - 1}}{e^{i{k_{ - 1,z}}z}}}&{z < 0},\\
				{{L_0}{e^{ - i{k_{0,z}}(z - z')}} + {M_0}{e^{i{k_{0,z}}(z - z')}}}&{z \ge 0}.
		\end{array}} \right.
	\end{equation}
Substituting Eq. \ref{eqn:>G} and Eq. \ref{eqn:<G} into Eq. \ref{eqn:Gp}, we obtain ${k_{0( - 1),z}} = \sqrt {\frac{{{\omega ^2}{\rho_{0(-1)} ^0}(z)}}{{C_{44,0(-1)}^0(z)}} - \frac{{C_{66,0(-1)}^0(z)}}{{C_{44,0(-1)}^0(z)}}\beta^2}$. 

Since the Green's function $\mathcal{G}$ and its normal derivative $C_{44}\partial_z\mathcal{G}$ are continuous at the slab-substrate interface and $C_{44}\partial_z\mathcal{G}$ is zero at the free surface $z=h$, we have 
	\begin{gather}
		{L_0}{e^{i{k_{0,z}}z'}} + {M_0}{e^{ - i{k_{0,z}}z'}} = {N_{ - 1}},\\
		i{k_{ - 1,z}}C_{44, - 1}^0{N_{ - 1}} = C_{44,0}^0\left( { - i{k_{0,z}}{L_0}{e^{i{k_{0,z}}z'}} + i{k_{0,z}}{M_0}{e^{ - i{k_{0,z}}z'}}} \right),
	\end{gather}
and
	\begin{equation}
		- i{k_{0,z}}{J_0}{e^{ - i{k_{0,z}}(h - z')}} + i{k_{0,z}}{K_0}{e^{i{k_{0,z}}(h - z')}} = 0.
	\end{equation}
	
At $z=z'$, the Green's function should still be continuous, leading to 
\begin{equation}
	{J_0} + {K_0} = {L_0} + {M_0}.
\end{equation}
In addition, integrating Eq. \ref{eqn:Gp} from $z=z'+0^-$ to  $z=z'+0^+$ leads to 
	\begin{equation}
		C_{44,0}^0\left( { - i{k_{0,z}}{L_0} + i{k_{0,z}}{M_0} + i{k_{0,z}}{J_0} - i{k_{0,z}}{K_0}} \right) = 1.
	\end{equation}
Solving these equations, we obtain
	\begin{gather}
	{M_0} = \frac{i}{{2{k_{0,z}}C_{44,0}^0}}\frac{{{e^{i{k_{0,z}}h}}}}{{{\eta _{ - 1}} - {e^{i{k_{0,z}}h}}}}\left( {1 + {e^{2i{k_{0,z}}(z' - h)}}} \right),\\
	{L_0} = \frac{i}{{2{k_{0,z}}C_{44,0}^0}}\frac{{{\eta _{ - 1}}}}{{{\eta _{ - 1}} - {e^{i{k_{0,z}}h}}}}\left( {1 + {e^{2i{k_{0,z}}(h - z')}}} \right),
		\end{gather}
	where\[{\eta _{ - 1}} = \frac{{{k_{0,z}}C_{44,0}^0 - {k_{ - 1,z}}C_{44, - 1}^0}}{{{k_{0,z}}C_{44,0}^0 + {k_{ - 1,z}}C_{44, - 1}^0}}{e^{ - i{k_{0,z}}h}}.\]
	We now have the full Green's function for $0\leq z<z'$ as following
	\begin{equation}
		\mathcal{G}(z,z') = \frac{i}{{2{k_{0,z}}C_{44,0}^0}}\frac{1}{{{\eta _{ - 1}} - {e^{i{k_{0,z}}h}}}}\left( {{\eta _{ - 1}}{e^{ - i{k_{0,z}}(z - z')}} + {\eta _{ - 1}}{e^{2i{k_{0,z}}h}}{e^{ - i{k_{0,z}}(z + z')}} + {e^{ - i{k_{0,z}}h}}{e^{i{k_{0,z}}(z + z')}} + {e^{i{k_{0,z}}h}}{e^{i{k_{0,z}}(z - z')}}} \right).
	\end{equation}
	Especially, for $z=0$,
	\begin{equation}\label{Green0}
		\begin{aligned}
			\mathcal{G}(0,z') = \frac{i}{{2{k_{0,z}}C_{44,0}^0}}\frac{{{\eta _{ - 1}}{e^{i{k_{0,z}}h}} + 1}}{{{\eta _{ - 1}} - {e^{i{k_{0,z}}h}}}}\left( {{e^{ - i{k_{0,z}}h}}{e^{i{k_{0,z}}z'}} + {e^{i{k_{0,z}}h}}{e^{ - i{k_{0,z}}z'}}} \right).
		\end{aligned}
	\end{equation}
	
\subsection{Radiation amplitude and accidental BIC}
We can now calculate the zeroth-order Fourier component of the Bloch mode at the $\Gamma$ point, which corresponds to the radiation amplitude. It is obained by solving Eq. \ref{eqn:eigenp} using the Green's function,
\be\label{sicwt}
Q_y^0({0^ - }) = \sum\limits_{G'=\pm\frac{2\pi}{a}}\int_{{0 }}^h {\mathcal{G}({0 },z') {\left[ {{\omega ^2}\rho_0 ^{ - G'} + {\partial _z}(C_{44,0}^{ - G'}{\partial _z})} \right]Q_y^{G'}(z')} dz'}.
\ee
Insert the solution of Eqs. \ref{ansatz} and \ref{Green0}, we obtain 
	\begin{equation}
		Q_y^0({0^ - })\sim C_{44,0}^{ - G'}{\kappa _0}\sin {\kappa _0}h\cos {k_{0,z}}h + \left( {{\omega ^2}\rho _0^{ - G'}(z') - C_{44,0}^{ - G'}(z')\kappa _0^2} \right)\frac{{{k_{0,z}}\sin {k_{0,z}}h\cos {\kappa _0}h - {\kappa _0}\sin {\kappa _0}h\cos {k_{0,z}}h}}{{k_{0,z}^2 - \kappa _0^2}},
	\end{equation}
	where the scaling factor in Eq. \ref{Green0} has been ignored and we only calculate the contribution from $G'=\frac{2\pi}{a}$ in order to identify the accidental BIC. The contribution from $G'=-\frac{2\pi}{a}$ is identical but with an opposite sign.
	
It is easy to see that $Q_y^0({0^ - })=0$ is achieved when
	\begin{equation}
	\label{eqn:FINAL}
	C_{44,0}^{ - G'}{\kappa _0}{k_{0,z}}\left( {{k_{0,z}}\tan {\kappa _0}h - {\kappa _0}\tan {k_{0,z}}h} \right) = {\omega ^2}\rho _0^{ - G'}(z')\left( {{\kappa _0}\tan {\kappa _0}h - {k_{0,z}}\tan {k_{0,z}}h} \right),
	\end{equation}
which corresponding to an accidental BIC at the $\Gamma$ point.

\begin{figure}
	\centering
	\includegraphics[width=0.8\linewidth]{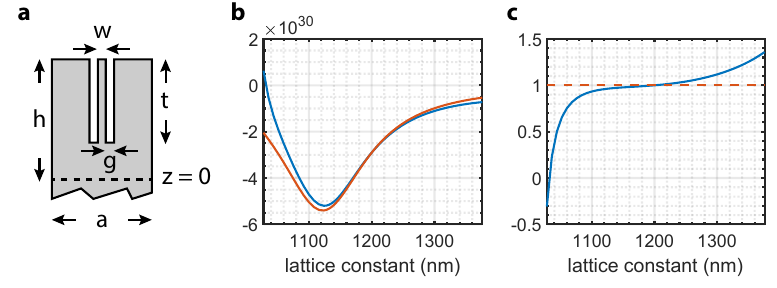}
	\caption{ \textbf{a}. Unit cell of the grating.  \textbf{b}. The left-hand side (blue) and right-hand side (red) of Eq. \ref{eqn:FINAL}. \textbf{c}. The ratio of left-hand side and right-hand side of Eq. \ref{eqn:FINAL}. Accidental BICs correspond to ratio equal 1.}
	\label{fig:figs1v1}
\end{figure}

We plot the L.H.S. and R.H.S. of Eq. \ref{eqn:FINAL} in Fig. \ref{fig:figs1v1}a and their ratio in Fig. \ref{fig:figs1v1}b for a grating with $g=100\,\mathrm{nm}$,  $t=1000\,\mathrm{nm}$, $w=120\,\mathrm{nm}$, and $h=1656\,\mathrm{nm}$. We find an accidental BICs can be realized for $a\approx1200\,\mathrm{nm}$.

\section{Temporal coupled-mode theory for IDT-coupled phononic crystal gratings}
\begin{figure}
		\centering
		\includegraphics[width=1\linewidth]{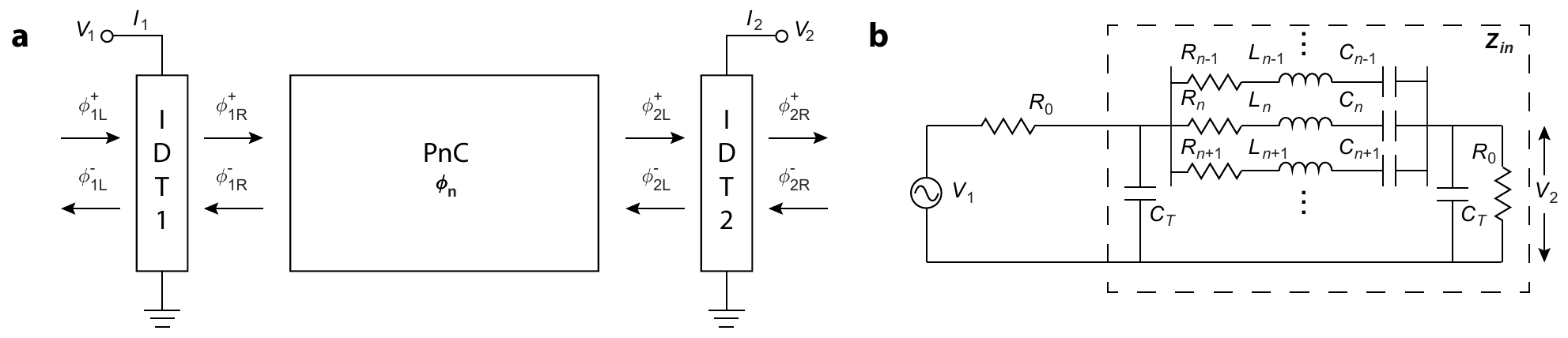}
		\caption{\textbf{a}. Schematic of the IDT-coupled phononic crystal grating. \textbf{b}. Equivalent circuit for the device probed by the network analyzer.}
		\label{fig:TCMT}
	\end{figure}

The transmission spectrum of the IDT-coupled phononic crystal grating can be modeled using the temporal coupled-mode theory. Consider the setup shown in the schematic of Fig. \ref{fig:TCMT}a, the temporal coupled-mode equations for the surface acoustic wave is given by
\begin{equation}
		\frac{d}{{dt}}\left| \bm{\phi}  \right\rangle  = (j\bm{\Omega}  - \bm{\Gamma} )\left| \bm{\phi}  \right\rangle  + \bm{K} \left| {{\bm{\phi} _{in}}} \right\rangle
\end{equation}
	and
	\begin{equation}
		\left| {{\bm{\phi} _ {out} }} \right\rangle  = \bm{T}\left| {{\bm{\phi} _ {in} }} \right\rangle  + \bm{D} \left| \bm{\phi}  \right\rangle,
	\end{equation}
where $\left| \bm{\phi}  \right\rangle  = {\left( {\begin{array}{*{20}{c}}
				{{\phi _1}}&{{\phi _2}}& \cdots &{{\phi _n}}
		\end{array}} \right)^T}$ are the amplitudes of phononic crystal resonances, $\bm{\Omega}$ and $\bm{\Gamma}$ represent their frequencies and dissipation rates, the matrix $\bm{K}$ ($\bm{D}$) denotes the coupling between the incoming (outgoing) surface acoustic waves $\left| \bm{\phi}_{in}  \right\rangle  = {({\begin{array}{*{20}{c}}
		{{\phi _{1R}^+}}&{{\phi _{2L}^-}}
	\end{array}})^T}$ ($\left| \bm{\phi}_{out}  \right\rangle  = {({\begin{array}{*{20}{c}}
	{{\phi _{1R}^-}}&{{\phi _{2L}^+}}
\end{array}})^T}$), and $\bm{T}$ is the direct transmission coefficient. For our experiment, $\phi _{1L}^+ = \phi _{2R}^-=0$ as the surface acoustic wave is generated by IDTs. IDTs' voltage and current responses are given by ${\phi_{1R}^+} =  \mu_1V_1$ and ${I_2} =  - {g_{m2}}\phi _{2L}^ -$, where $\mu_1$ ($g_{m2}$) is the transmitter (receiver) response function. 

The relation between $I_2$ and $V_1$ can be obtained by solving the temporal coupled-mode equation, which is given by
 \begin{equation}
	\label{eqn:curr}
	{I_{2}} = G\left[ \sum\limits_n {\frac{{{\kappa _{n,e}}}}{{j(\omega  - \omega_n) + {\kappa _n}/2}}} \right]{V_1},
\end{equation}	
where $G=-\mu_1g_{m2}$. Thus the admittance $Y_{21}$ of the phononic crystal device is
\begin{equation}
	\label{eqn:adm}
	{Y_{21}} = I_2/V_1=\sum\limits_n {\frac{{G{\kappa _{n,e}}}}{{j(\omega  - \omega_n) + {\kappa _n}/2}}}.
\end{equation}
Comparing to the admittance of $RLC$ series resonators in parallel,
\begin{equation}
	\label{eqn:RLC}
	Y = \sum\limits_n {\frac{1}{{{R_n} + j\omega {L_n} + 1/j\omega {C_n}}}}  = \sum\limits_n {\frac{1}{{{R_n} + 2j{L_n}(\omega  - {\omega _n})}}},
\end{equation}
where ${\omega _n} = \sqrt {1/{L_n}{C_n}}$, we find the phononic crystal grating with multiple resonances can be modeled as $RLC$ series resonators in parallel, with the equivalent circuit components: ${L_n} = 1/2G{\kappa _{n,e}}$, ${R_n} = {\kappa _n}/2G{\kappa _{n,e}}$ and ${C_n} = 2G{\kappa _{n,e}}/\omega _n^2$. 

Fig. \ref{fig:TCMT}b shows the full equivalent circuit corresponding to the device probed by a network analyzer, where $C_T$ is the static capacitance of the IDT and  the reference impedance of the port is $R_0 = 50\,\Omega$. The transmission spectrum can be derived as \cite{steer2019microwave}
	\begin{equation}
		|{S_{21}}|^2 \equiv \left| \frac{V_2}{V_1} \right|^2 = {\left| {\frac{{2{R_0}}}{{{Z_{in}} + {R_0}}}} \right|^2},
	\end{equation}
where $Z_{in}$ is the total input impedence to the network analyzer as shown in Fig. \ref{fig:TCMT}b. We assume the system satisfies $GR_0\ll1$ (weak piezoelectricity), $\omega C_TR_0\ll1$, and the resonances are well-resolved, i.e., $\kappa_n<|\omega_{n+1}-\omega_n|$, then we have
\begin{equation}
	\label{eqn:TCMTSpec}
	|{S_{21}}{|^2} \approx {\left| {\frac{{2{R_0}}}{{1/{Y_{21}} + 2{R_0}}}} \right|^2} = {\left| {\frac{{2{R_0}\sum\limits_n {\frac{{G{\kappa _{n,e}}}}{{j(\omega  - {\omega _n}) + {\kappa _n}/2}}} }}{{1 + 2{R_0}\sum\limits_n {\frac{{G{\kappa _{n,e}}}}{{j(\omega  - {\omega _n}) + {\kappa _n}/2}}} }}} \right|^2} \approx \sum\limits_n {{{\left| {\frac{{2G{R_0}{\kappa _{n,e}}}}{{j({\omega _n} - \omega ) + {\kappa _n}/2}}} \right|}^2}}
\end{equation}
We used Eq. \ref{eqn:TCMTSpec} to fit the measured transmission spectrum.
	
\section{Finite-size effect on $Q$}\label{sec:finite}
\begin{figure}[H]
	\centering
	\includegraphics[width=0.8\linewidth]{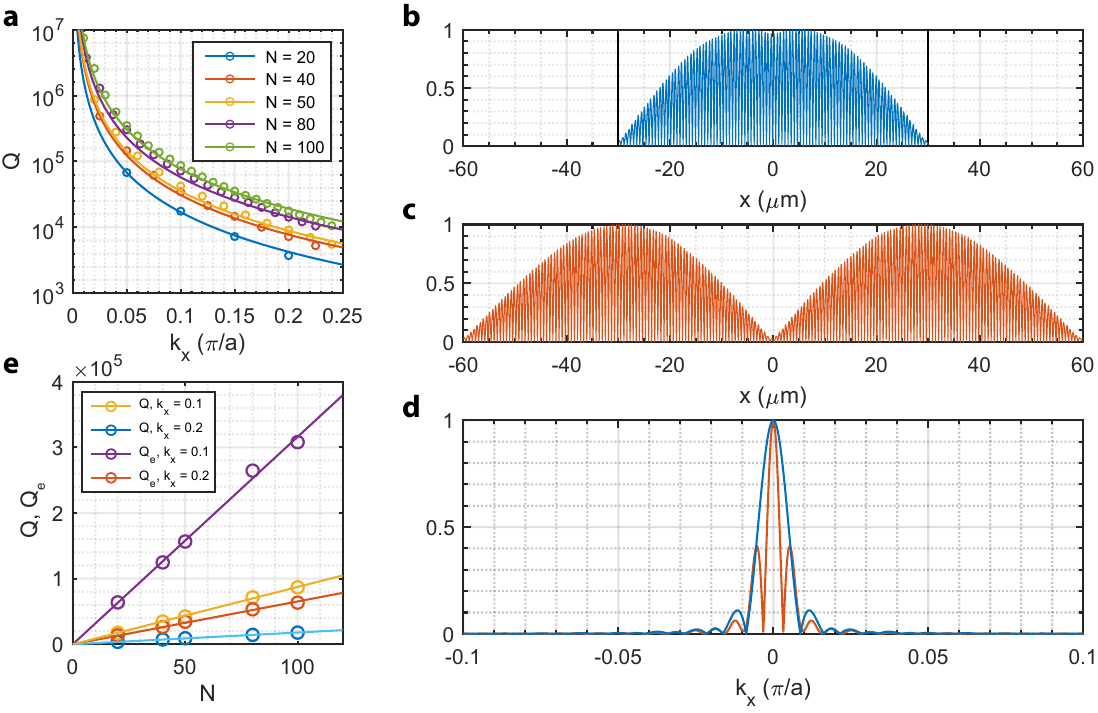}
	\caption{\textbf{a}. Radiative quality factor of standing-wave resonances characterized by $k_x$ in gratings with size $N$. \textbf{b} and \textbf{c}. Mode envelope of the first-order resonance of the $N=50$ grating (\textbf{b}) and the second-order resonance of the $N = 100$ grating (\textbf{c}). \textbf{d}. Normalized Fourier transform of the mode envelope function of \textbf{b} (blue) and \textbf{c} (red). \textbf{e}. $Q$ and $Q_e$ dependence on $N$. }
	\label{fig:Finite}
\end{figure}

The finite-size phononic crystal grating not only discretizes the band structure but also affects the envelope function of the standing-wave resonances \cite{akahane2005fine,liu2022optomechanical}. We simulated $Q$ of the standing-wave resonances in gratings with $N = 20$,  40, 50, 80 and 100, which is shown in Fig. \ref{fig:Finite}a. The $m-$th order standing-wave resonance is denoted by the wavevector $k_x = m\pi/Na$. We find $Q$ of standing-wave resonances of different-size gratings are not the same even though they have the same $k_x$. This can be explained by the ``squeezing" of the mode envelope function by the finite grating boundary, which causes momentum spreading in the $k-$space. To illustrate this, Figs. \ref{fig:Finite}b and c show the mode envelope of the first-order resonance of the $N=50$ grating and the second-order resonance of the $N = 100$ grating, respectively. Fig. \ref{fig:Finite}d shows the $k-$distribution of the two resonances obtained by Fourier transform of the mode envelope function. It is seen that the first-order resonance of the $N=50$ grating occupies more $k$ components far from the $\Gamma$ point, thus leading to a lower quality factor. We also find the $Q$ and $Q_e$ of standing-wave resonances with the same $k_x$ is linearly proportional to $N$ (Fig. \ref{fig:Finite}e). This might be explained by a general relation $Q_{k_x}(N/N_c)$, where $N_c$ characterizes the size of grating where $Q_{k_x}$ approaches the quality factor of the Bloch mode with $k_x$ of the unit cell. Thus, for $N/N_c\ll 1$, $Q_{k_x}(N/N_c)\approx \alpha\frac{N}{N_c}$ by Taylor expansion and keeping the leading-order term, which explains the linear dependence on $N$. Similarly it is true for $Q_e$.

\section{Analysis of measured $Q$}
The $Q$ of the standing-wave resonance of a fabricated finite phononic crystal grating can be decomposed as following
\begin{equation}
	\frac{1}{Q} = \frac{1}{{{Q_r}}} + \frac{1}{{{Q_s}}}+ \frac{1}{{{Q_e}}} + \frac{1}{{{Q_a}}},
\end{equation}
where $1/Q_r$ is the radiation loss due to the finite size, $1/Q_s$ is the scattering loss due to the disorder, $1/Q_e$ is the coupling loss to the surface acoustic wave, and $1/Q_a$ is the material absorption loss. As shown in Section \ref{sec:finite}, $Q_r$ is proportional to $N$. Scattering loss is caused by disorder-induced intermodal scattering and subsequent radiation, and thus $Q_s$ is also proportional to $N$. The external quality factor $Q_e$ is also proportional to $N$ as shown in the previous section. As a result, we write
\begin{equation}\label{QN}
	\frac{1}{Q} =  \frac{A}{N}\left( {\frac{1}{{{Q_{r,0}}}} + \frac{1}{{{Q_{s,0}}}} + \frac{1}{{{Q_{e,0}}}}} \right) + \frac{1}{{{Q_a}}},
\end{equation}
where $Q_{r,0}$, $Q_{s,0}$, $Q_{e,0}$ and $Q_a$ are size-independent quality factors.

The wavevector $k_x$ of the observed lowest-order resonances in the phononic crystal grating can be determined by the quality factor of resonances. We analyze gratings with lattice constant far from the merging point, such that the scattering loss is not suppressed by the merging BIC. We designate the wavevector of the observed lowest-order resonance as $k_{x0}$. To determine $k_{x0}$, we perform a linear regression using $1/NQ_{r,0}$ as the independent variable, where $Q_{r,0}$ is the simulated unit-cell radiation quality factor, and measured $1/Q$ as the dependent variable for the observed standing-wave resonances. The slope obtained from the linear regression is the parameter $A$ in Eq. \ref{QN} and the intercept $B$ can be expressed as: 
\begin{equation}
	\label{eqn:intercept}
	B = \frac{A}{N}(\frac{1}{{{Q_s}}} + \frac{1}{{{Q_{e,0}}}}) + \frac{1}{{{Q_a}}}.
\end{equation}

\begin{figure}
	\centering
	\includegraphics[width=0.8\linewidth]{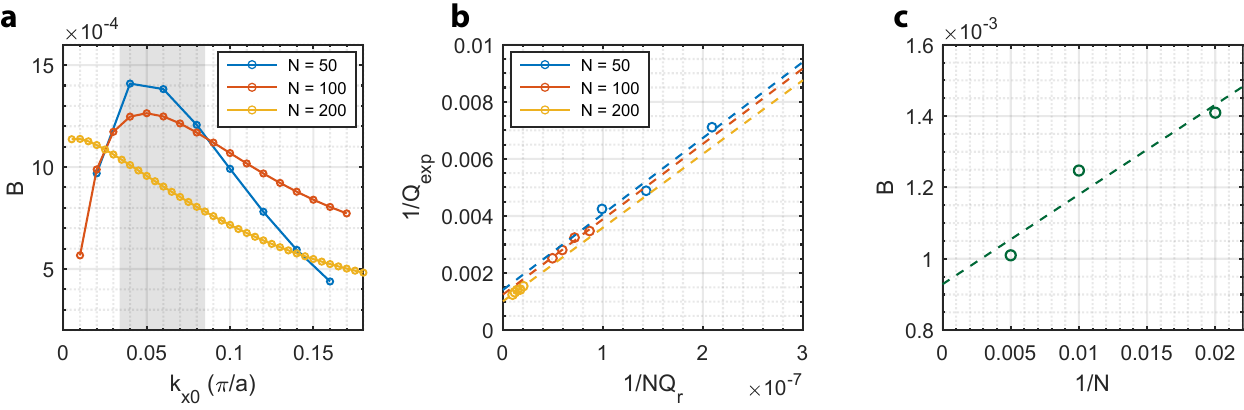}
	\caption{\textbf{a}. Intercept of the linear regression for different trial $k_{x0}$. \textbf{b}. Linear regression results for $k_{x0} = 0.04\pi/a$. \textbf{c}. Intercepts for different sizes $N$ with $k_{x0} = 0.04\pi/a$. }
	\label{fig:figs4}
\end{figure}

For different trial values of $k_{x0}$, the independent variable $1/NQ_r$ varies, resulting in different $A$ and $B$. From Eq. \ref{eqn:intercept}, we can see that the regression intercept $B$ should increase for decreasing grating size. We plot  $B$ versus trial $k_{x0}$ for the three grating sizes in Fig. \ref{fig:figs4}a. Only for $k_{x0}$ in the range of $0.04\pi/a-0.08\pi/a$, $B$ satisfies such a relation. The fitting result corresponding to $k_{x0}=0.04\pi/a$ is shown in Figs. \ref{fig:figs4}b and c.

%

	%